\documentclass[11pt,letterpaper]{article}

\makeatletter
\def\@xfootnote[#1]{%
  \protected@xdef\@thefnmark{#1}%
  \@footnotemark\@footnotetext}
\makeatother

\pdfoutput=1
\usepackage{amsmath,amssymb,calc}
\usepackage{graphicx}
\usepackage{float}
\usepackage{amstext}    
\usepackage{array}    
\usepackage[utf8]{inputenc} 
\usepackage{color}
\usepackage{chngcntr}
\usepackage[autostyle]{csquotes}
\usepackage{hyperref}

\counterwithin*{equation}{section}

\renewcommand{\theequation}{\thesection.\arabic{equation}}
\usepackage{comment}

\newcommand\be{\begin{equation}}
\newcommand\ee{\end{equation}}
\newcommand{\bea}{\begin{eqnarray}}
\newcommand{\eea}{\end{eqnarray}}

\newcommand{\nn}{\nonumber}

\def\id{\protect{{1 \kern-.28em {\rm l}}}}

\def\D{\Delta}

\def\k{\kappa}

\def\1{^{(1)}}
\def\0{^{(0)}}
\def\2{^{(2)}}

\def\id{\protect{{1 \kern-.28em {\rm l}}}}

\newcommand*{\affaddr}[1]{#1} 
\newcommand*{\affmark}[1][*]{\textsuperscript{#1}}

\def\Cincy{Dept. of Physics, University of Cincinnati,
2600 Clifton Ave, Cincinnati, OH 45221, USA}
\def\Fermilab{Fermi National Accelerator Laboratory, P.O. Box 500, Batavia, IL 60510, USA}

\begin{document}

\date{}
\title{\Large\bfseries Decay of Ultralight Axion Condensates}

%

\author{%
Joshua Eby,\affmark[1,2,]\!
	\footnote{Electronic Address: \it{joshaeby@gmail.com}} \,
Michael Ma,\affmark[1,]\!
	\footnote{Electronic Address: \it{mam@ucmail.uc.edu}}\,
Peter Suranyi,\affmark[1,]\!
	\footnote{Electronic Address: \it{peter.suranyi@uc.edu}} \,
and 
L.C.R. Wijewardhana\affmark[1,]\!
	\footnote{Electronic Address: \it{rohana.wijewardhana@uc.edu}}
\vspace{.3cm} \\
{\it\affaddr{\affmark[1]\Cincy}}\\
{\it\affaddr{\affmark[2]\Fermilab}} 
}

\begin{titlepage}
\maketitle
\thispagestyle{empty}

\vskip 2 cm

\begin{abstract}
Axion particles can form macroscopic condensates, whose size can be galactic in scale for models with very small axion masses $m\sim10^{-22}$ eV, and which are sometimes referred to under the name of Fuzzy Dark Matter. Many analyses of these condensates are done in the non-interacting limit, due to the weakness of the self-interaction coupling of axions. We investigate here how certain results change upon inclusion of these interactions, finding a decreased maximum mass and a modified mass-radius relationship. Further, these condensates are, in general, unstable to decay through number-changing interactions. We analyze the stability of galaxy-sized condensates of axion-like particles, and sketch the parameter space of stable configurations as a function of a binding energy parameter. We find a strong lower bound on the size of Fuzzy Dark Matter condensates which are stable to decay, with lifetimes longer than the age of the universe.
\end{abstract}

\end{titlepage}

\section{Introduction}
Axions are hypothetical particles introduced to solve the Strong $CP$ problem in QCD \cite{PQ1,PQ2,Weinberg,Wilczek,DFS,Zhitnitsky,Kim,Shifman}.  Axions are self-adjoint bosons, with no conserved discrete quantum numbers to guarantee particle number conservation. The axion potential can be written in terms of an angular variable with a $2\pi$ shift symmetry. Axion-like scalar particles also appear in a variety of models beyond QCD, especially in low-energy limits of string theories \cite{Dimopoulos,ChoiKim,Dine}. Those axions have properties similar to QCD axions, but their mass scales and decay constants are vastly different. 

Axion-like particles are among the prime candidates for the composition of dark matter~\cite{Preskill,Sikivie1,Davidson,DF,Holman,Sikivie2}.  Axions, being scalar bosons, can condense. Axion condensates have been discussed by numerous groups, with condensate sizes ranging from galaxy or galaxy cluster scale \cite{SikivieYang,Witten}, down to stellar size and smaller (termed ``axion stars'') \cite{KolbTkachev,Iwazaki,BarrancoNS,TkachevFRB,ESVW,Guth}, all the way to radii of a few meters \cite{BB,Braaten}. Also of interest is the cosmological evolution of the axion field, which has been worked on extensively in \cite{Khlopov1,Khlopov2,Khlopov3,Khlopov4,Khlopov5}, but this will not be disussed further here.

Uncondensed QCD axions are not stable, as they can decay to photons through a process $a\to 2\,\gamma$, but the decay rate is slow enough such that most axions created after the Big Bang would survive many Hubble times~\cite{TkachevPossibility}. Axion condensates, however, may also decay slowly due to the self-interaction of axions \cite{ESW,Braaten2016}. The self-interaction term of their Lagrangian (for both the instanton~\cite{nonchiral,Vecchia} and chiral~\cite{Vecchia,Cortona} cases) have only terms containing an even number of axion fields. Thus, disregarding the rare decay into photons, the axion number is conserved only modulo 2. 

In a recent paper~\cite{ESW} we have investigated the decay of weakly bound axion stars due to the self-interaction of axions.  The decay proceeds mostly through a sequence of processes,
\be\label{theprocess}
{\cal{A}}_N\to{ \cal{A}}_{N-3}+a_p,
\ee
where ${\cal{A}}_k$ is an axion star, which is a condensate containing $k$ axions, and $a_p$ denotes an axion in a scattering state with the magnitude of the momentum $p$.
 The process (\ref{theprocess}) is the simplest of many possible decay modes responsible for the decay of axion stars.\footnote{The decay rates via processes ${\cal{A}}_N\to{ \cal{A}}_{N-5}+a_p$ or ${\cal{A}}_N\to{ \cal{A}}_{N-4}+a_{p_1}+a_{p_2}$ are significantly lower and unlikely to have any cosmological significance \cite{ESW}.}  This process is allowed by energy-momentum conservation, provided the binding energy of a bound axion is small enough that a relativistic particle can be produced: $\delta E < 2\,m\,/\,3$, where $m$ is the mass of a free axion.
 
In~\cite{ESW} we used an axion field operator, which was the generalization of the field proposed by Ruffini and Bonazzola~\cite{RB}. To facilitate the decay process, terms creating and annihilating axions in the continuum of scattering states were included  in the quantum field of axions, in addition to  terms creating and annihilating bound axions~\cite{RB}.  The Ruffini-Bonazzola method is based on taking the expectation value of the quantum Einstein and Klein-Gordon equations in the condensate to derive equations of motion for the metric components and the scalar field.
We solved the equations of motions numerically in the  weak gravity and weak binding ($\delta E\ll m$) limits, to find the wave function of axions in the condensate~\cite{ESVW}. 

The bound axions {\em are not in momentum eigenstates}. They have definite energies, but their wave functions extend over the size of the axion star.  Accordingly, the bound axions have an extended momentum distribution as well. In the recent publications~\cite{Braaten2016,Braaten2}, the authors questioned the validity of the decay mechanism proposed in~\cite{ESW}, arguing that momentum is not conserved in (\ref{theprocess}), and that the decay rate through this process is exactly zero by the Optical Theorem. These authors have further suggested that one can show the rate to be zero by the classical equation of motion for the condensate. We will address these issues and explain our response in Appendix A.

In the present paper we will apply our method of discussing the decay of dilute axion stars \cite{ESW} to condensates of cosmological size; such models have been referred to previously as Fuzzy Dark Matter (FDM) \cite{Hu}. Condensates of galactic sizes have been considered by a number of authors, and typically correspond to a scalar particle mass of $m\sim 10^{-22}-10^{-21}$ eV \cite{Turner,Press,Sin,Hu,Goodman,Peebles,Amendola,Shapiro,Schive,Marsh,Witten}.\footnote{For a brief but recent review of ultralight scalar field dark matter models, see \cite{Lee} and references therein.} When such condensates are formed from real scalars, a version of the decay analysis of \cite{ESW} applies, and we will investigate whether interesting bounds can be placed on these models by taking decays into account. As we will explain in the next section, we will utilize the axion potential with a cosine dependence on the field; other proposals, for example a $\cosh$ potential \cite{Sahni,Alcubierre}, have been investigated in the context of ultralight scalars as well.

It is also an aim of this paper to emphasize the inclusion of axion self-interactions in investigations of axion condensates. Although the self coupling $\lambda \sim m^2/f^2 \sim 10^{-95} \lll 1$ for typical FDM models ($f$ is the axion decay constant), the astronomical number of axions in a condensate participating in these interactions could lead to large corrections to certain important physical quanties. We investigate the macroscopic properties of these condensates using the fully self-interacting analysis and emphasize the differences from the non-interacting limit.

In the next section, we give a more detailed explanation of how axion star decay through the process (\ref{theprocess}) can be calculated. In Section \ref{WavefunctionSec}, we will outline the calculation of the wavefunction, following largely \cite{ESVW}. In Section \ref{DecayRateSec}, we apply the formulas for the macroscopic properties and decay rates to condensates formed from ultralight axion-like particles. We conclude in Section \ref{ConclusionSec}.

We will use natural units throughout, where $\hbar = c = 1$.

 
\section{Decay through self interaction} \label{OverviewSec}
There is a variety of methods for the  quantitative investigation of axion condensates, as classical~\cite{TkachevPossibility,Iwazaki,KolbTkachev,Kouvaris2,Lee2}, quantum mechanical~\cite{ChavanisMR,ChavanisMR2,Guth,Kouvaris,Guzman1,Guzman2,Guzman3}, and field theoretic~\cite{BB,ESVW,Braaten}. Our field theoretic discussion of the decay of an axion condensates into relativistic axions \cite{ESW} was based on an extension of the Ruffini-Bonazzola operator~\cite{RB}, by the addition of scattering state contributions.\footnote{Appendix B we will discuss why a continuous spectrum of scattering state solutions can be added to the boson field operator. Furthermore, we will also discuss why using free spherical wave scattering states is quite sufficient in our calculations.} Thus, we proposed to extend the expansion of the boson field using the form \cite{ESW}
\be\label{field2}
\Phi (r,t) = R(r)\,e^{-i\,E_0\,t}\,a_0 + R(r)\,e^{i\,E_0\,t}\,a_0^\dagger 
     + \psi_f(r,t) + \psi_f^\dagger(r,t),
\ee
where $E_0$ is the energy eigenvalue of a single bound axion, and where $R(r)$ and $a_0$ are the wave function and annihilation operator of the axions in the condensate (respectively). $\psi_f(r,t)$ is the annihilation part of a complete system of free axion operators expanded in scattering states,
\be \label{free}
\psi_f(r,t) =\frac{1}{2\,\pi^2}\sum_{l,m}Y_{l}^m(\hat{r})
	  \int_0^\infty \frac{dp\,p}{2\,\omega_p}\, 
	    j_l(p\,r)\,e^{-i\,\omega_p\,t}a_{lm}(p),
\ee
where $\omega_p$ and $a_{lm}(p)$ are the energy eigenvalue and the annihilation operator of the scattering state axion, with quantum numbers $l$ and $m$, respectively. The functions $j_l(x)$ and $Y_{l}^m(\hat{x})$ are spherical Bessel functions and spherical harmonics, respectively. 

The  annihilation operator in the spherical wave basis is defined by
\be
a_{lm}(p)= i^l\, p\int d\Omega_p\,Y_{l}^{m*}(\hat{p})\,a (\vec p),
\ee
where $a(\vec p)$ is the annihilation operator for a particle which is the eigenstate of the momentum operator with eigenvalue $\vec p$ (i.e. a plane wave). This annihilation operator and its adjoint, the creation operator, satisfy the commutation relation 
\be
[a_{lm}(p),a_{l'm'}^\dagger(p')] =
	2\,\omega_p\, (2\,\pi)^3 \,\delta(p-p')\,\delta_{ll'}\delta_{mm'}.
\ee
Note that (\ref{free}) is exactly equal  to the negative frequency part of a complete system of free axion states given by
\be
\frac{1}{(2\,\pi)^3}\int\frac{d^3p}{2\,\omega_p}
      e^{i(\vec{p}\cdot\vec{r}-\omega_p\,t)}a(\vec p),
\ee
which was used in~\cite{ESW} to investigate the decay of QCD axion stars.
 
We will see later that the bosons produced by the decay of a weakly bound boson condensate are relativistic. Therefore, we have chosen to use free particle states to approximate the scattering states. For the purposes of this calculation, this choice is admissible, because the energy level of produced axions is sufficiently high compared to the effective depth of the potential created by gravitation and self-interactions. This is explained in greater detail in Appendix B. In a future work we will take into account corrections to this approximation.

Let us consider now (\ref{theprocess}) in the Born approximation. The axion self-interaction potential can be approximated by the so-called instanton potential \cite{nonchiral,Vecchia}
\be \label{instanton}
 V(\Phi) = m^2\,f^2\,\Big[1 - \cos\Big(\frac{\Phi}{f}\Big)\Big],
\ee
where $m$ and $f$ are the axion mass and decay constant (respectively). In the Ruffini-Bonazzola paradigm, one finds the expectation value of eq. (\ref{instanton}) transforms the cosine into a Bessel function $J_0$ \cite{ESVW}.\footnote{More generally, the annihilation process for $k$ bound axions generates an effective potential proportional to the Bessel function $J_k$, as explained in the Appendix of \cite{ESW}.} One also finds that the transition matrix element for the process (\ref{theprocess}) is
\begin{align}\label{matrix1}
{\cal M}_3 &= \int dt\,d^3r\,\langle N\vert \,V(\Phi) \vert N-3,\,\phi(p)\rangle \nonumber \\
      &= -i\, m^2\,f\,\int dt\,dr\,r^2\, J_3\left[X(r)\right]\,
	      e^{3\,i\,E_0\,t} \int d\Omega_r\langle 0\vert\psi_f(r ,t)\vert \phi(p)\rangle
\end{align}
where $X(r)=2\,\sqrt{N}\,R(r)\,/\,f$ is the rescaled wave function of the condensate which, as we will see in Section \ref{WavefunctionSec}, can be obtained by solving the equations of motion~\cite{ESVW}. We are considering transitions of the form (\ref{theprocess}), where $\langle N \vert$ is the initial $N$ particle condensate (the left hand side of (\ref{theprocess})), and $\vert N-3,\,\phi(p)\rangle$ is the direct product of the final state $N-3$ particle condensate and a scattering state $\phi(p)$ of momentum magnitude $p$ (the right hand side of (\ref{theprocess})). 

We restrict this work to non-rotating axion condensates; the reason for this is twofold. First, for the potential in eq. (\ref{instanton}) and the parameters we use here, only the inner cores of galaxies composed of axion particles can be described as a condensate; outside of this inner region, the dark matter halo is described by a virialized gas of particles \cite{Witten} and cannot participate in the decay processes we describe here.\footnote{For ultralight bosons with \emph{repulsive} self-interaction, like those presented in e.g. \cite{Goodman,Shapiro}, the condensates can be much larger, and can even constitute the entire dark matter halo.} Because the condensed core is small compared to the full radius of the halo, it carries at most a tiny fraction of the angular momentum of the galaxy, so as a first approximation we believe restricting to $\ell=0$ angular momentum states is appropriate. The second reason is that a full treatment of rotating axion condensates has not yet been done, though slowly rotating condensates were analyzed in a particular limit by \cite{Davidson}. This is a topic we hope to return to in the near future.

Because we work in the limit of zero angular momentum, annihilation processes of the form $a\,a\to G$, where $G$ is a spin-2 graviton \cite{SuperRad1,SuperRad2}, have a rate of zero. Such an interaction would require the participating axions to have at least $\hbar$ of angular momentum each; and even if we accounted for the nonzero rotation speed of the galaxy, by our estimation the vast majority of axions in a galactic condensate would have far less angular momentum than what would be necessary for this process to occur.

For static condensates, note that the integration over $\Omega_r$ in eq. (\ref{matrix1}) vanishes for all but $s$-wave axions.\footnote{Should we consider rotating axion stars, higher angular momentum scattering states would also contribute. } In that case, the scattering state axion is described by the zero angular momentum contribution only, $\vert \phi(p)\rangle = a_{00}^\dagger(p)\vert 0\rangle$. Then the wave function of  the emitted axion is
\be
\phi(r,t)=\langle 0\vert \psi_f(r ,t)\vert \phi(p)\rangle
	  =\sqrt{4\,\pi}\,\frac{e^{-i\,\omega_p\,t}\sin(p\,r)}{r}.
\ee
The integration over $t$ also fixes the energy of the outgoing axion to $\omega_p=3\,E_0$. The matrix element takes the form~\cite{ESW}
\be\label{matrixr}
{\cal M}_3 =-4\,\pi^2\sqrt{4\pi}\,f\,
	\delta(3\,E_0-\omega_p)\,I_3(p),
\ee
where the dimensionless integral is
\begin{align} \label{I3}
 I_3(p) 
      &= m^2\,\int_{-\infty}^{\infty} dr\,r\,e^{i\,p\,r}\,J_3\left[X(r)\right] \nn \\
      &\approx \frac{m^2}{48}\,\int_{-\infty}^{\infty} dr\,r\,e^{i\,p\,r}\,X(r)^3.
\end{align}
The symmetry of the integrand for the substitution $r\to-r$ has been used to extend the integration region to all real values of $r$, and to switch from $\sin(p\,r)$ to $e^{i\,p\,r}$ in the integrand. In the second equality, we expanded the Bessel function $J_3$ to leading order, an appropriate limit for dilute axion stars.

Now observe that for dilute axion stars the radius of the star $R$ is very large. In other words, $X(r)$ has a  large coordinate uncertainty, $\delta r\sim R \sim (m\,\Delta)^{-1}$, where $\Delta=\sqrt{1-(E_0\,/\,m)^2}\ll1$ \cite{ESVW,ESW}.  As a result,  the range of $p$, as represented by the momentum uncertainty $\delta p\sim m\,\Delta \ll m$, is very small.  Then due to the delta function in eq. (\ref{matrixr}), enforcing energy conservation, the emitted axion has a momentum peaked at a very large value, $p\simeq \sqrt{8}\,m$; as a result, the matrix element (\ref{matrixr}) is very small for  weak binding. However, as the binding energy $\delta E$ increases, ${\cal M}_3$ will take larger values and the decay rate $\Gamma\sim |{\cal M}_3|^2$ also increases. 

Now to bring out issues related to momentum conservation, we can define the momentum representation wave function as
\be\label{fourier}
\Xi(q)=\frac{1}{(2\,\pi)^3}\int d^3r\,X(r)\,e^{i\, \vec{q}\cdot\vec{r}}.
\ee
Then we can rewrite eq. (\ref{matrixr}) as
\be\label{matrixp}
{\cal M}_3=- \frac{\pi^2\,\sqrt{4\,\pi}\,m^2\,f}{12}\,\delta(3\,E_0-\omega_p)\,
	\int \delta^3(\vec{p}-\vec{q}_1-\vec{q}_2-\vec{q}_3)
	    \prod_{k=1}^3\,\Xi(q_k)\,d^3q_k.
\ee
Since for weakly bound condensates $p\simeq \sqrt{8}\,m$, the magnitude of the transition matrix depends crucially on the large $q$ tail of momentum distribution $\Xi(q)$.  However, for calculational purposes, it is still advantageous to use (\ref{matrixr}) rather than (\ref{matrixp}). One can rely on the numerical solution of the equations of motion, as explained below, using a simple approach to estimate approximate behavior of $\Xi(q)$ at large $q$~\cite{ESW}.

Note that the process (\ref{theprocess}) is not the only possible channel through which decay can proceed; however, it is by far the dominant process. First, we have shown previously \cite{ESW} that processes of the form ${\cal A}_N\to {\cal A}_{N-(2\,j+1)}+a_p,$ are suppressed by higher powers of $\Delta$ for each higher $j>1$. In the weak-binding limit, where $\Delta \ll 1$, these corrections are completely negligible. On the other hand, this argument breaks down for dense axion stars \cite{Braaten,ELSW}, where $\Delta = \mathcal{O}(1)$; we will return to this case in a future publication.

Second, there are processes of the form $ {\cal A}_N\to {\cal A}_{N-k}+\mu\,a_p,$ where $\mu > 1$ axions are emitted at once. Unlike the process (\ref{theprocess}), the emission of $\mu>1$ axions from a condensate can proceed on-shell, implying that the corresponding decay rate has a weak dependence on $\Delta$. Nonetheless, as shown in \cite{ESW}, these processes are suppressed by the very small factor $m^2/f^2$ for each additional axion in the final state. Since in FDM $m^2/f^2 \sim 10^{-95} \lll 1$, we can safely neglect these corrections as well. We conclude that (\ref{theprocess}) is by far the dominant contribution to the decay of axion condensates.

\section{The calculation of the wave function of the condensate} \label{WavefunctionSec}
We review here the calculation of the condensate wavefunction $X(r)$ \cite{ESVW}. The matrix elements of the $rr$ and $tt$ components of the Einstein equation, along with the Klein-Gordon equation, form a closed set of equations for the metric and the axion field, $X(r)$:
\bea
\frac{A'}{A}&=&\frac{1-A}{r}+2\,\pi\,r\,\delta\,A\left\{\frac{E_0{}^2\,X^2}{B}
	  +\frac{X'^2}{A}+m^2X^2-\frac{m^2}{16}X^4\right\},\label{Ett}\\
\frac{B'}{B}&=&\frac{A-1}{r}+2\,\pi\,r\,\delta\,A\left\{\frac{E_0{}^2 \, X^2}{B}
	  +\frac{X'^2}{A}-m^2X^2+\frac{m^2}{16}X^4\right\},\label{Err}\\
X''&=&-\left[\frac{2}{r}+\frac{B'}{2\,B}-\frac{A'}{2\,A}\right]X'
	  -A\left[\frac{E_0{}^2\,X}{B}-m^2X+\frac{m^2}{8}\,X^3\right]\label{KG},
\eea
where the metric is 
\be \label{metric}
ds^2=-B(r)\,dt^2+A(r)\,dr^2+r^2\,d\Omega^2,
\ee
with $\delta = f^2/M_P{}^2$ and $M_P =1\,/\,\sqrt{G}= 1.22\times10^{19}$ GeV (the Planck mass). As above, we have taken only the leading contribution to the Bessel function which represents the self-interaction potential; doing so preserves the leading attractive $X(r)^4$ interaction term.

Assuming that $\delta\ll1$, a condition satisfied in applications where gravity is weak (Newtonian limit), we can write $A = 1 + \delta \,a$ and $B = 1 + \delta\,b$, where $a,\,b=\mathcal{O}(1)$. Furthermore using the large radius approximation and the definition of the scale parameter $\Delta=\sqrt{m^2-E_0{}^2}/m$, we can introduce dimensionless radial coordinate as $x= m\,r\,\Delta$. In that case the axion field also scales with its engineering dimension, such that $X(r)=\Delta\,Y(x)$, leading to the following system of equations for $a$, $b$, and $Y$ in leading order of $\Delta$ and $\delta$~\cite{ESVW}:
\bea\label{eomY}
Y''(x)&=&[1+\kappa \,b(x)]Y(x)-\frac{2}{x}\,Y'(x)-\frac{1}{8}\,Y(x)^3,\nn\\
a'(x)&=&\frac{x}{2}\,Y(x)^2-\frac{1}{x}\,a(x),\nn\\
b'(x)&=&\frac{1}{x}\,a(x),
\eea
where\footnote{This definition of $\kappa$ appears to differ by a factor of $8\pi$ compared with \cite{ESW}, because in that work we wrote $\delta$ in terms of the reduced Planck mass. In fact, the two definitions of $\k$ are equivalent.} $\kappa=8\pi\delta\,/\Delta^2$. Since $b(x)$ is proportional to the Newtonian gravitational potential, $\kappa\sim G$ is the effective coupling constant of the field $Y(x)$ to gravity. Further details, and a more comprehensive justification of this double expansion of the equations, can be found in \cite{ESVW}.

Solutions of the equations of motion (\ref{eomY}) correspond to ground-state configurations of axions, which can be stable or metastable. In \cite{ESVW}, we solved these equations and found a spectrum of solutions which were parameterized by $\kappa$ (or, equivalently, by $\Delta$). When applied to QCD axion parameters $m=10^{-5}$ eV and $f=6\times10^{11}$ GeV, we found a maximum gravitationally stable mass of $M_c \sim 10^{19}$ kg.\footnote{The maximum masses for attractive interactions were discussed by Stoof \cite{Stoof2} in the context of condensed matter BECs, and in the context of boson stars by Chavanis and Delfini \cite{ChavanisMR,ChavanisMR2}.} By rescaling these solutions to values of $m$ and $f$ appropriate for FDM, we can analyze the properties of galactic-size condensates in a way that includes the self-interaction term in the potential. The physical interpretation of these condensates is that they form the cores of FDM halos; they are surrounded by a virialized distribution of axions which extend to the outer edge of the dark matter halo.

To analyze the decay of these condensates through processes like (\ref{theprocess}), we investigate the singularity structure of solutions of (\ref{eomY}). Now, (\ref{eomY}) is a system of equations with two singular points, $x=0$ and $x=\infty$. Using boundary conditions we require that the solutions are regular at $x=0$ and decrease exponentially at $x\to\infty$. The solutions are even functions of $x$, so they also approach zero at $x\to -\infty$. Then they are real analytic functions at $-\infty<x<\infty$ and can be continued into the complex plane of $x$. As they fast vanish at infinity, the contour of integration can be moved up along the imaginary axis in the rescaled version of the integrals in eqs. (\ref{matrixr}) and (\ref{I3}), until we reach a singularity in the complex plane.  The contribution of that singularity dominates the decay rate integrals at large momentum. 

It is easy to show that the Klein-Gordon equation (\ref{eomY}), in which the the leading order singular terms are retained, is\footnote{In fact, this expression contains the next to leading order term proportional to $Y'(x)$.}
\be
Y''(x)+\frac{2}{x}Y'(x)+\frac{1}{8}Y(x)^3 = 0.
\ee
 Near the singular point $x=i\,\rho$, this has a solution of the form
\be\label{singular}
Y(x)=\frac{8\,\rho}{x^2+\rho^2}-\frac{2}{3\,\rho}-\frac{1}{18\,\rho^3}(x^2+\rho^2)+\mathcal{O}([x^2+\rho^2]^2).
\ee
The parameter $\rho$ is an integration constant, having a one-to-one relationship with the rescaled central density of the axion field $Y(0)^2$, and in turn, with mass and the radius of the condensate. In fact, high order Taylor series expansion of equations around $x=0$ show that the singularity closest to the origin is indeed of the form (\ref{singular}), connecting the value of $Y(0)$ with $\rho$ \cite{ESW}. In principle, gravitational interactions have an effect on the solutions $Y(x)$. In practice, however, the term in the equations of motion (\ref{eomY}) which couple $Y(x)$ to gravity give a subleading contribution to the singularity. We can therefore solve the equations in the limit $\kappa\ll 1$, i.e. where gravity decouples. In that limit, the nontrivial solution has $Y(0)=12.268$, which implies a fixed value $\rho=.603156$.

Finally, we can rewrite the Fourier transform of eq. (\ref{fourier}) as
\be
\Xi(q) = \frac{1}{(2\,\pi)^2\,i\,q\,m^2\,\Delta}\int_{-\infty}^\infty dx\, x \,\exp\Big(\frac{i\,q\,x}{m\,\Delta}\Big)\,Y(x).
\ee
In this form, it is clear that at small $\Delta$ the singular term of (\ref{singular}) term dominates the integral. 
To calculate the decay rate, we follow the procedure of \cite{ESW}: we take the leading order solution for $Y(x)$ near the singularity $x = i\,\rho$, given by eq. (\ref{singular}), and evaluate  $I_3(p)$ in the matrix element of eq. (\ref{matrixr}). The result is
\be
I_3(p_0)\simeq\frac{32\,i\,\pi}{3}\frac{\rho}{\Delta}
    \exp\left(-\frac{2\sqrt{2}\,\rho}{\Delta}\right),
\ee
where $p_0=\sqrt{9\,E_0{}^2-m^2}\simeq \sqrt{8}\,m$.

The decay rate for the process (\ref{theprocess}) is then
\begin{equation}
 \Gamma_3 =\frac{1}{T} \int\,\frac{dp}{(2\pi)^32\,\omega_p}\,
	  \Big|\mathcal{M}_3\Big|^2
    = \frac{2\,\pi\,f^2}{p_0}\Big|I_3(p_0)\Big|^2,
\end{equation}
where $T$ is the duration of the decay process. Then the lifetime of the condensate through this decay process is
\begin{equation} \label{dtaudN}
 \frac{d\tau}{dN} \simeq m\,\frac{d\tau}{dM}
	\simeq -\frac{1}{3\,\Gamma_3}.
\end{equation}
Further details regarding the evaluation of eq. (\ref{dtaudN}) can be found in \cite{ESW}; the result for the process (\ref{theprocess}) is
\begin{equation} \label{lifetime}
 \tau = \frac{3\,y_M\,\Delta^2}{4096\,\pi^3\,\rho^3\,m}
	\exp\left(\frac{4\,\sqrt{2}\,\rho}{\Delta}\right),
\end{equation}
where $y_M\simeq 75.4$ is determined by the relationship between $M$ and $\Delta$ in the large $\Delta$ region \cite{ESVW}. The lifetime is a monotonically decreasing function of $\Delta$ in the relevant range; in the case of QCD axions, we found in \cite{ESW} that above a value $\Delta\simeq .05-.06$, axion condensates become very unstable to decay to relativistic axions, their lifetimes becoming shorter than the age of the universe. We will examine the consequences of this fact in the context of ultralight axions in the next section.

\section{Stable spectrum of ultralight axion condensates} \label{DecayRateSec}
Very light axion fields can have de Broglie wavelengths as large as entire dark matter halos, possibly implying a connection between these two scales. Ultralight bosons appear often in theories of physics beyond the Standard Model, including those requiring compactification of extra dimensions. Such models, termed ``Fuzzy Dark Matter'' (FDM) \cite{Hu}, have been written about extensively \cite{Turner,Press,Sin,Hu,Goodman,Peebles,Amendola,Shapiro,Schive,Marsh,Witten}. While there are significant constraints on these models\footnote{While this work was being finalized, a paper appeared suggesting a strong tension between the preferred mass scale for FDM, $m\sim 10^{-22}-10^{-21}$ eV, and data from Lyman-$\alpha$ forest simulations \cite{Irsic}. We will not comment here about whether such constraints could rule out FDM as a viable paradigm.}, they remain a viable alternative to WIMP or QCD axion models of dark matter. 

\begin{figure}[t]
 \includegraphics[scale=1]{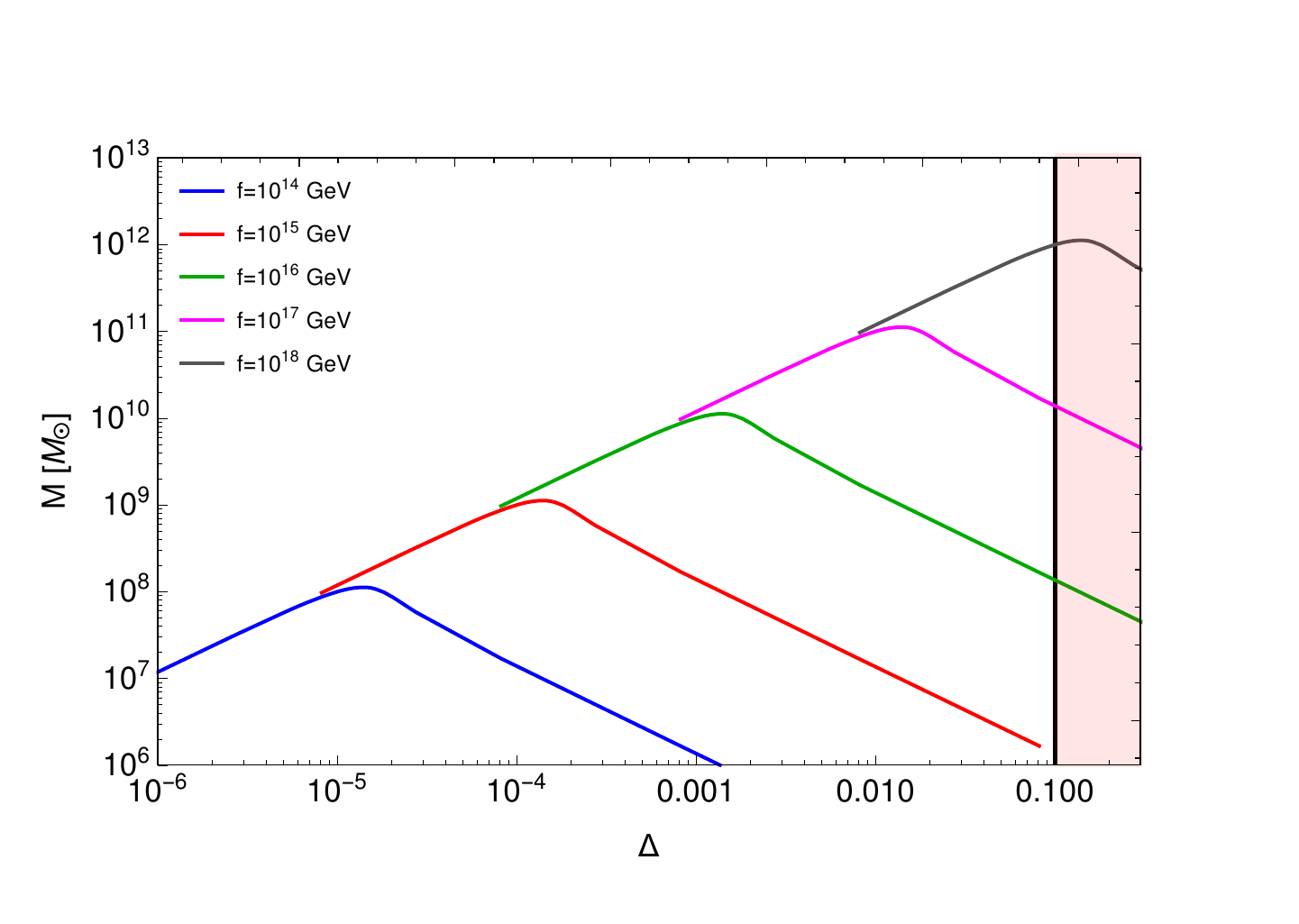}
 \caption{The allowed masses for condensates of axion particles in FDM, as a function of the binding energy parameter $\Delta$; these condensates constitute the cores of FDM halos. Axion condensates in the shaded region are unstable to decay to relativistic axions with a very short lifetime. Here we have used the model parameters $m=10^{-22}$ eV, and $f$ in the range between $10^{14}$ and $10^{18}$ GeV; increasing the particle mass $m$ merely shifts these curves down proportionally to $1/m$.}
 \label{Mass_FDM}
\end{figure}

\begin{figure}[t]
 \includegraphics[scale=1]{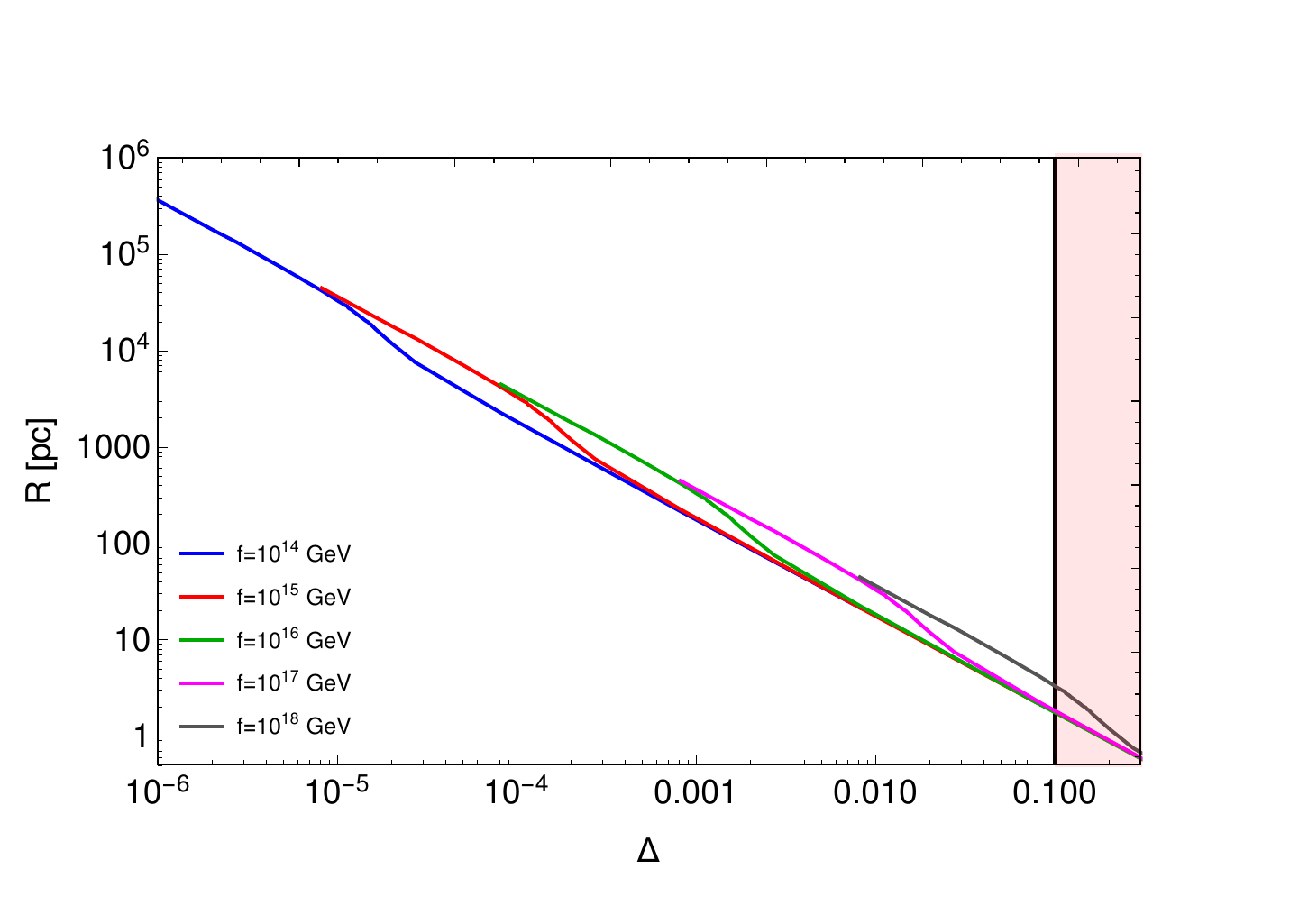}
 \caption{The allowed radii for condensates of axion particles in FDM, as a function of the binding energy parameter $\Delta$; these condensates constitute the cores of FDM halos. Axion condensates in the shaded region are unstable to decay to relativistic axions with a very short lifetime. Here we have used the model parameters $m=10^{-22}$ eV, and $f$ in the range between $10^{14}$ and $10^{18}$ GeV; increasing the particle mass $m$ merely shifts these curves down proportionally to $1/m$.}
 \label{Radius_FDM}
\end{figure}

We will consider an ultralight axion in this context, using the potential of eq. (\ref{instanton}). The mass of the ultralight axion in question will be taken to be $m \sim 10^{-22}$ eV, which gives the right approximate scale for the size of dark matter halos \cite{Witten},
\begin{equation}
 \frac{\lambda_{dB}}{2\pi} = \frac{1}{m\,v} 
 = 1.92 \text{ kpc}\Big(\frac{10^{-22}\text{ eV}}{m}\Big)
		  \Big(\frac{10\text{ km/sec}}{v}\Big),
\end{equation}
where $v$ is the velocity in the halo. This choice is also consistent with the known epoch of matter-radiation equality. Further, a decay constant of $f \sim 5\times10^{16}$ GeV naturally leads to the correct relic density, and can thus account for the observed dark matter abundance \cite{Witten}; however, to remain as general as possible, we allow $f$ to deviate from this value by a few orders of magnitude. At the upper limit of what we consider, $f=10^{18}$ GeV is still below the Planck scale, implying that the parameter $\delta = f^2/M_P^2 \approx .007 \ll 1$; thus, the weak-gravity approximation holds reasonably well over our entire range.

In \cite{ESVW}, we found numerically the solutions to the system (\ref{eomY}) over a wide range of $\kappa$. In the FDM picture, these solutions correspond to the  cores of FDM halos discussed (most recently) in \cite{Witten}. We found that there exists a maximum mass at $\kappa \approx .34$, above which axion condensates are gravitationally unstable. On the stable branch of masses $\kappa >.34$, the mass $M$ and radius\footnote{We use the common convention that $R_{99}$, the radius inside which $.99$ of the mass of the condensate is located, represents the ``size'' of the condensate.} $R_{99}$ of the condensate  are fit by the functions \cite{ESVW}
\begin{equation} \label{MandR}
 M(\kappa) \approx \frac{8.75}{\sqrt{\kappa}}\,\frac{M_P\,f}{m}, \qquad
	R_{99}(\kappa) \approx 1.15\,\sqrt{\kappa}\,\frac{M_P}{f\,m}.
\end{equation}
We observe in Figure \ref{Mass_FDM} that at fixed $m$, the value of $f$ determines the position of the maximum mass, and thus the turnaround of the function $M(\Delta)$. A similar structure can be observed in Figure \ref{Radius_FDM} for the radius, where the position of the maximum mass corresponds to a slight dip in the otherwise straight line representing $R_{99}(\Delta)$. Trading $\kappa$ for $\Delta$ in eq. (\ref{MandR}), we see that the lines
\begin{equation}
 M(\Delta) \approx 1.75\,\Delta\,\frac{M_P{}^2}{m} \qquad
	R_{99}(\Delta) \approx \frac{5.75}{m\,\Delta}
\end{equation}
bound the full set of solutions from above.

For the specific choice of FDM parameters $m=10^{-22}$ eV and $f=5\times10^{16}$ GeV we find the maximum mass  $M_c \approx 6\times10^{10} M_\odot$; this is lower than the value found in the non-interacting limit of $8\times10^{11} M_\odot$ \cite{Witten} by about an order of magnitude, due to the inclusion of attractive self-interactions. Our estimate of the maximum mass also agrees well with the recent analysis of \cite{MarshUL}, which also includes the leading attractive self-interaction. More generally, the mass and radius of FDM halo cores over a wide range of the scale parameter $\D$ and at different values of $f$ are illustrated in Figures \ref{Mass_FDM} and \ref{Radius_FDM}.

We can also analyze the relationship between $M$ and $R_{99}$, which were investigated for both attractive and repulsive self-interactions in \cite{ChavanisMR,ChavanisMR2}. In a recent paper \cite{Witten}, the authors present a bound on the product
\begin{equation} \label{MR_NI}
 M\,R_{1/2} \geq 3.925\,\frac{M_P{}^2}{m^2} \qquad \text{(non-interacting bosons)},
\end{equation}
where $R_{1/2}$ is the radius inside which $.5$ of the mass of the condensate is contained; the inequality is saturated for stationary, ground state configurations, i.e. for the condensates considered here. In our calculation, on the stable branch of solutions (where $\kappa > .34$), we find the product
\begin{equation}
 M\,R_{99} = 10.06\,\frac{M_P{}^2}{m^2},
\end{equation}
using eq. (\ref{MandR}). To find the relationship between $R_{99}$ and $R_{1/2}$, we calculated their ratio numerically and found that $R_{99}/R_{1/2} \approx 3.65$ holds within 1\%, in a range of $1/\Delta$ extending over many orders of magnitude. This implies
\begin{equation}
 M\,R_{1/2} = 2.76\,\frac{M_P{}^2}{m^2}  \qquad \text{(interacting axions)}.
\end{equation}
This product is below the lower bound (\ref{MR_NI}) presented in \cite{Witten} due to our inclusion of self-interactions.

It is worth noting also that in the limit $f \sim M_P$, the mass-radius relationship for axion condensates approaches that of a black hole. This is easy to see using eq. (\ref{MandR}):
\begin{equation}
 \frac{G\,M}{R_{99}} = \frac{1}{M_P{}^2}\,
	    \frac{\frac{8.75}{\sqrt{\kappa}}\frac{M_P\,f}{m}}
	      {1.15\sqrt{\kappa}\frac{M_P}{f\,m}}
	  = \frac{7.6}{\kappa}\,\frac{f^2}{M_P{}^2}.
\end{equation}
Near $\kappa=\mathcal{O}(1)$ (the position of the maximum mass) and $f\sim M_P$, we find $G\,M/R_{99}\sim 1$, implying that $R_{99} \sim R_S$, the Schwarzschild radius.

We must also ensure that the weak-binding approximation, on which our analysis \cite{ESW} and the classical one of \cite{Witten} depends, is also valid.\footnote{A stability analysis for very strongly-bound condensates, with $\Delta=\mathcal{O}(1)$, is a task we plan to undertake in the near future.} We observe in Figure \ref{Radius_FDM} that cores of radius $R\lesssim 1$ pc have $\Delta \gtrsim .3$, and become relatively strongly bound. Such cores would not be well-described by our weak-binding analysis.
 
An estimate of the decay rate, obtained from the expression derived in \cite{ESW}, is given in eq. (\ref{lifetime}); it is a one-to-one function of the binding energy parameter $\Delta$ in the region of interest. Because the condensate mass $M$ is determined by the value of $\Delta$, it is easy to connect $\tau$ to $M$ as well. Following the analysis of \cite{ESW}, we find that axion condensates with $m\sim 10^{-22}$ eV which have $\Delta \gtrsim .1$ have lifetimes shorter than the age of the universe. This region is represented by the shaded regions in Figures \ref{Mass_FDM} and \ref{Radius_FDM}. For $f\lesssim 10^{18}$ GeV, the transition to decay instability occurs on the gravitationally unstable branch of solutions; however at $f\gtrsim 10^{18}$ GeV, the bounds from decay are as strong or stronger than those coming from gravitational stability. This can be an important constraint on bound structures originating in theories of Planck scale axions.

In Figure \ref{Radius_FDM}, it is particularly striking that almost regardless of the value of $f$, condensates with $R\lesssim 2$ pc lie in the unstable, shaded region. This implies a fundamental limiting radius of $R_{min}\sim 2$ pc for FDM cores composed of axions. Observe also that, in Figure \ref{Mass_FDM}, it is easy to read off the maximum mass of FDM condensates for each value of the decay constant $f$. For any axion theory with $f\ll M_P$, no stable condensate exists with a mass larger than about $M_{max} \approx 10^{12} M_\odot$.

 \section{Conclusions} \label{ConclusionSec}
In a previous publication~\cite{ESVW} we established scaling relations for the mass and radius of weakly bound condensates of interacting axions, as functions of the mass of the axion, its decay constant, and the particle energy (or alternatively the central density). We also found the maximum mass and size of the bound states as functions of those parameters.  In this paper we have applied those results to condensates of axions forming FDM, providing corrections to similar calculations which neglect the self-interaction of axions \cite{Sin,Hu,Goodman,Amendola,Schive,Marsh,Witten}.

In another publication~\cite{ESW} we developed methods to calculate the lifetime of axion condensates due to their self-interaction, through the four-axion interaction term in which three bound axions produce a single free relativistic axion. Here we have applied those results  to estimate the lifetime of condensates formed from FDM. We have found that, provided the decay constant of FDM axions satisfies $f\lesssim 0.05 \,M_P$, all condensates having  binding energy smaller than that of those of maximal mass have lifetimes greater than the age of the universe making them viable candidates for forming central regions of galactic halos. We have also explained in details the decay mechanism described in \cite{ESW} and further clarified the justification of its validity.

The methods we have described here, based on previous work in \cite{ESVW,ESW}, rely on a double expansion to leading order in $\delta$ and $\Delta$. This is appropriate for so-called dilute axion stars, which are weakly bound. However, it has been pointed out that an effective short-distance repulsive interaction in the axion potential also gives rise to dense axion stars \cite{Braaten,ELSW,ELSW2}, which are at least energetically stable. We plan to extend our methods to this regime to analyze the properties of these states in the near future.

Collapsing boson condensates have been investigated by a number of groups \cite{Stoof,Stoof2,Harko,ChavanisCollapse,ELSW,Tkachev2016,MarshCollapse,ELSW2}. Recently, we found that supercritical QCD axion condensates, having masses larger than the maximal allowed stable mass, collapse towards the global minimum of the effective axion potential~\cite{ELSW,ELSW2}. Similar arguments indicate that FDM axion condensates which exceed the maximum mass $M_c$ will also collapse in this way. Such supercritical condensates can form during galactic collisions, in a manner similar to the mechanism outlined in \cite{Collisions}; such events would lead to collapse, causing the condensate to emit a large number of relativistic particles. Consequences of such events  will be studied in a future publication.

 \section*{Acknowledgements}
 We thank M. Amin, P. Argyres, E. Braaten, J. Brod, P. Fox, R. Harnik, A. Kagan, A. Mohapatra, K. Schutz, G. Semenoff, M. Takimoto, H. Zhang, and J. Zupan for fruitful discussions.  The work of J.E. was supported by the U.S. Department of Energy, Office of Science, Office of Workforce Development for Teachers and Scientists, Office of Science Graduate Student Research (SCGSR) program. The SCGSR program is administered by the Oak Ridge Institute for Science and Education for the DOE under contract number DE-SC0014664. J.E. also thanks the Fermilab Theory Group and the Weizmann Institute Department of Physics for their hospitality. L.C.R.W. thanks Mainz Institute for Theoretical Physics for their hospitality, and the participants of the Quantum Vacuum and Gravitation program, especially M. Bartelmann, A. Mazumdar, and T. Prokopec, for discussions.

 \appendix
 \renewcommand{\theequation}{A.\arabic{equation}}
 \section*{Appendix A: The ${\cal A}_N\to {\cal A}_{N-3}+a_p$ decay}
 The results of our paper, "The Lifetime of Axion Stars" \cite{ESW}, have been called into question in recent publications~\cite{Braaten2016, Braaten2}. Before discussing this issue in detail, we review the premises of our work.  We proposed a way to discuss the decay of axion stars through the repeated elementary decay mode
 \be\label{theprocess2}
 {\cal A}_N\to {\cal A}_{N-3}+a_p,
 \ee
 where ${\cal A}_k$ denotes an axion star, which is a condensate of $k$ axions and $a_p$ denotes a (relativistic) free axion labeled by its momentum $p$.  This calculation was performed using an extended axion field operator, eq. (\ref{field2}), which included both bound and scattering state contributions.
 
 First of all, we need to establish the fact that there is no conservation law that would forbid (\ref{theprocess2}). Axions are real bosons, and consequently the axion number is not conserved.  Axions, being coupled to the electromagnetic field as $\Phi\,E\cdot B$, can decay to photons through the slow process $a\to 2\,\gamma$, though this decay process does not significantly affect the lifetime of axion stars~\cite{TkachevPossibility}. Disregarding axion decay to photons, the axion number is conserved modulo 2, as the self-interaction terms of the axion potential contains only even powers of the axion field, through a dependence of $\cos(\Phi\,/\,f)$ in eq. (\ref{instanton}).  Momentum and energy conservation would allow the decay process to proceed even if the condensates were in momentum eigenstates:  a decay process is always allowed if the sum of the masses of the decay products is smaller than the mass of the decaying object, unless the conservation of discrete quantum numbers prevents it. 
 
 At any rate, the condensates represented in (\ref{theprocess2}) are not in momentum eigenstates. They are quantum objects, which have extended wave functions localized on a large radius $R$. Consequently, though they have mean momentum of zero, their momentum distributions are smeared.  In fact, the momentum distributions extend from zero to infinity, albeit with fast decreasing amplitudes. 
 
 We have to emphasize that a Gross-Pit\"aevskii approach, being the non-relativistic limit of the relativistic quantum field theory, cannot be used to discuss particle number violating processes. In the non-relativistic limit the axion number is conserved and process (\ref{theprocess2}) is forbidden. Note however that a method to adapt the Gross-Pit\"aevskii approach to these processes was formulated by the authors of \cite{MTY}. We also point out Gell-Mann's totalitarian principle~\cite{GM}, borrowed from  T. H. White's "The Once and Future King", which when it is applied to physics is: "Everything not forbidden is compulsory." This principle implies that the process (\ref{theprocess2}) should exist. In the present context, this principle also implies that taking the nonrelativistic limit discards important contributions that make such a transition possible.
 
Now the questions raised in~\cite{Braaten2016,Braaten2} have two sides: (a) microscopic, and (b) macroscopic.  The microscopic argument (a) pertains to the question of momentum conservation in the elementary process of three bound axions at rest turn into a moving axion.  The macroscopic argument (b) concerns the whole condensate, namely how momentum is conserved in the overall process (\ref{theprocess2}), i.e. how the momentum carried away by the produced scattering state axion is transferred to the outgoing axion star, ${\cal A}_{N-3}$.    

The argument of \cite{Braaten2016,Braaten2} pertaining to the microscopic side (a) is that in the overwhelmingly dominating Born approximation, the process around which the decay  process (\ref{theprocess}) is built is
\be
3\,a_c \to a_f
\ee
where $a_c$ represents a bound axion and $a_f$ represents the final state axion, is inadmissible due to energy-momentum conservation. The total energy of the three bound axions is $E_{tot}\simeq 3\,m$, which is certainly sufficient to produce a free relativistic axion.  However, if the three axions were in momentum representation, having zero momentum in the rest system of the axion star, then the axion in the final state would not have the required momentum of $p\simeq \sqrt{8} m$.

To resolve  this problem, observe that the three axions contributing to the decay process are not at rest.  The notion of "rest" is tied to particles in momentum representation, and for the axions in the axion star, this is not so.  They have momentum space wave functions, which, though peaked near zero, extend to large momenta (albeit with very small probability). As we discussed above, the uncertainty of the momentum of each axion is $\delta p \sim R^{-1}$. The probability that three of the bound axions have sufficient total momentum to create a free axion,  $p=\sqrt{9 \,E_0{}^2-m^2}$, decreases rapidly with the  size of a condensate, though it is {\em never exactly zero}. Consequently, local momentum conservation always allows the decay of condensate. The question of whether this decay process affects cosmology, due to the survival or non-survival of the condensate, is a question of numerical calculations and depends on the parameters of the axion theory, as well as the size and mass of the condensate.

Another argument presented in \cite{Braaten2016,Braaten2} to the local momentum conservation is based on the optical theorem. The argument invokes diagram
\be\label{optical}
3\,a_c\to a_f \to 3\,a_c
\ee
claiming that the propagator of the axion $a_f$ in the intermediate state does not have an imaginary part, because presumably its denominator, $E_{tot}^2-p^2-m^2\simeq 8 \,m^2\not=0$. The diagram having no imaginary part, the decay rate must vanish. However, this argument is based again on the premise that the momentum of the condensed axions in the initial and final states of the process (\ref{optical}) is zero. This is not a valid assumption, since those particles are not in momentum eigenstates and with a tiny probability they can produce sufficient momentum to allow the axion in the intermediate state to go on mass-shell. A comprehensive discussion of how the imaginary part appears in a process like (\ref{optical}) for particles in a condensate can be found in~\cite{MTY}.

The macroscopic argument (b) claims that even if the above discussed elementary process is possible, there is no mechanism by which the three axions participating in the decay process transmit the momentum to the axion star as a whole; so (the argument goes) contrary to the requirement of momentum conservation in (\ref{theprocess2}) the recoil of the axion star ${\cal A}_{N-3}$ in the final state of (\ref{theprocess2}) is not possible. 

Consider, however, that the condensate itself is  {\em not in momentum representation} either, unlike condensates in a container, for which the boundary conditions are set by the wall of the container. Still, one could evoke the valid argument that just like the expectation value of the coordinate, the expectation value of the momentum of the condensate must be zero. Consequently, the only constraint we can impose on the decay process is the conservation of the {\em expectation value of the} momentum.  Now, the condensate of the final state of (\ref{theprocess2}) is not in momentum representation either, though its average momentum is zero as well. Consider now the created scattering state axion.   It is produced as a zero angular momentum spherical wave, going with the same probability into every direction.   Though the magnitude of the momentum is sharply peaked at a particular value, the spherical wave also has a vanishing average  momentum.  Thus, the average momentum is conserved in (\ref{theprocess2}). 

One should not confuse the emission of the axion with its detection.  Suppose we detect an emitted axion, the decay product of (\ref{theprocess2}).  By performing a measurement we alter the system.  Just like in the case of the collapse of a simple wave packet by performing a measurement, the conservation of the average momentum is valid only if we include the measuring device, which absorbs the appropriate amount of momentum, to make the average momentum of the complete system zero.  

We turn now to the last critique presented in \cite{Braaten2016,Braaten2}, namely that the classical equation of motion for the axion star precludes a linear coupling to a scattering state axion. In \cite{Braaten2016}, the authors write about our previous work \cite{ESW} as follows:

\begin{quote}
 \emph{They expanded the scalar axion field $\phi$ around the classical field $\phi_0$: $\phi = \phi_0 + \tilde{\phi}$, where $\tilde{\phi}$ is the quantum fluctuation field. The Hamiltonian includes a term proportional to $\phi_0{}^3\,\tilde\phi$ from the axion interaction potential. They claimed that this term produces transitions of $3$ condensate axions into one relativistic axion of energy $3\,m_a$. However, the sum of all terms in the Hamiltonian that are linear in $\tilde\phi$ is zero by the classical equations of motion for $\phi_0$. Thus the matrix element for producing a single relativistic axion is $0$.}
\end{quote}

On the contrary, we have emphasized (and explain further in Appendix B) that a complete set of states satisfying the Klein-Gordon equation for the axion field contains both bound and scattering state components, $\Phi = \Phi_b + \Phi_s$. Our expansion does not distinguish a classical and a quantum component, and as such, one cannot identify $\Phi_b$ as $\phi_0$, a purely classical field. By direct calculation we see that the matrix element for (\ref{theprocess2}) is nonzero and is proportional to $\Phi_s\langle N |\Phi_b{}^3|N-3\rangle$.

We can see this more quantitatively by analyzing the equation of motion. The standard procedure to calculate the equation of motion is to take the matrix element 
\begin{equation}
 \langle N | KG[\Phi] |N-1\rangle = 0
\end{equation}
of the Klein-Gordon operator $KG[\Phi]$. This is precisely what we have used in eq. (\ref{KG}) and is equivalent to the Gross-Pitaevskii equation used by numerous other authors, e.g. \cite{Guth,Braaten,ChavanisMR}. In an ``exact'' field theory, any matrix element of the form $\langle\psi|KG[\Phi]|\psi'\rangle=0$, including
\begin{equation} \label{KGN-3}
 \langle N | KG[\Phi]|N-3\rangle=0.
\end{equation}
However, because this exact theory is not known, one is restricted to using some ansatz for the field $\Phi$. We have chosen the Ruffini-Bonazzola ansatz, expanded to include the scattering state solutions, which was given in eq. (\ref{field2}). With this choice, the wavefunction $R(r)$ (which is equivalent to the Gross-Pitaevskii wavefunction), does not satisfy eq. (\ref{KGN-3}), and in this same parameterization, the matrix element for the process (\ref{theprocess2}) is nonzero as well. 

We are working to extend the Ruffini-Bonazzola paradigm so that eq. (\ref{KGN-3}), as well as higher-order expressions in the operator Klein-Gordon equation, can be simultaneously satisfied. In this extension, we find that the rate for the process (\ref{theprocess2}) is still nonzero, and is equal at leading order to the result we obtain here. We will present these and related findings in a future publication.

 \renewcommand{\theequation}{B.\arabic{equation}}
  \section*{Appendix B: The  continuous spectrum in the Ruffini-Bonazzola equations} \label{AppB}
 
 The most general bound solution, discussed in~\cite{RB}, for a real scalar field in spherically symmetric metric of eq. (\ref{metric}) and satisfying the non-interacting Klein-Gordon equation is
 \be\label{bound}
 \Phi_b=  \sum_{nlm}c_{nlm}R_{nl}(r)\,Y_l^m(\theta,\phi)\,e^{-i\,E_{nl}\,t}+c.c.
 \ee
 where $c_{nlm}$ are arbitrary amplitudes of the $n$th bounds state solution with angular momentum $l$.  The wave functions $R_{nl}(r)$ satisfy wave equations
  \be\label{KG3}
  R_{nl}''=-\left(\frac{2}{r}+\frac{B'}{2\,B}-\frac{A'}{2\,A}\right)R_{nl}'
	  -A\left(\frac{E_{nl}{}^2}{B}-m^2-\frac{l(l+1)}{r^2}\right)R_{nl}.
	  \ee
To be able to describe condensates of bosons, Ruffini and Bonazzola \cite{RB} introduced second quantization by promoting coefficients $c_{nlm}$ to creation and annihilation operators $c_{nlm}\to a_{nlm}$ and $c_{nlm}^*\to a_{nlm}^\dagger$; these operators satisfy
\be
[a_{nlm},a_{n'l'm'}^\dagger]=\delta_{nn'}\delta_{ll'}\delta_{mm'}.
\ee

Now notice that $\Phi_b$ is not the most general solution of the wave equation.  Because the attractive gravitational potential, which makes bound states solutions possible, vanishes at large $r$, (\ref{KG3}) has scattering state solutions as well. After quantization these states can be written as
\be\label{scattered}
\Phi_s=\frac{1}{2\,\pi^2}\int \frac{d^3k}{2\,\omega_k} \sum_{ml}
      \left[ f_l(k)Y_l^m(\theta,\phi)\,e^{-i\,\omega_k\,t}a_{lm}(k)+h.c.\right],
\ee
where $k=\sqrt{\omega_k^2-m^2}$ is the momentum and
\be\label{commutator}
[a_{lm}(k),a_{l'm'}^\dagger(k')]=(2\,\pi)^3 2\,\omega_k\,\delta_{ll'}\delta_{mm'}\delta(k-k').
\ee
The complete set of solutions to the Klein-Gordon equation is $\Phi=\Phi_b+\Phi_s$.

The Ruffini-Bonazzola method was used by Barranco and Bernal~\cite{BB} and by us~\cite{ESVW} to describe axion stars.  It is sufficient to use of quantized field $\Phi_b$ in leading order to describe static axion stars.  However, the term $\Phi_s$ becomes significant if we notice that in first order of the expansion of the axion potential using $\Phi$ results in an operator 
\be
L_i=\frac{1}{24}\frac{m^2}{f^2}\Phi_b^3\Phi_s+ \text{ (terms higher order in $\frac{\Phi_b}{f}$)}.
\ee
This has been described in detail in~\cite{ESW} and also applied in the present paper.

The second subject we discuss in this Appendix is the reason why \emph{free} spherical waves, $f_l(k)\to j_l(k\,r)$, can be used to calculate the decay rate of axion stars.  This is the consequence of the fact that dilute axion stars, those we consider in this work, are weakly bound. In fact, only axion stars whose particle energy satisfies $|E_0 - m| \lesssim 0.002\,m$, corresponding to a value of $\Delta\approx.05$, could survive from the big bang until the present epoch.

To further justify our use of free spherical waves in the final state, consider (\ref{KG}) and (\ref{eomY}).  Using $Y(x)=X(r)/\Delta$ and $x=m\,r\,\Delta$, where $\Delta^2\simeq -2\,(E_0-m)/m$, the Klein-Gordon equation in (\ref{eomY}) can be rewritten as the nonlinear Schr{\"o}dinger equation
\begin{align}\label{KG4}
(E_0-m)\,X(r) &= -\frac{1}{2\,m}\left[ X''(r) + \frac{2}{r}X'(r)\right] 
      + \kappa\,(E_0 - m)\,b(r)\,X(r) + \frac{m}{16}X(r)^3\nonumber\\
    & \equiv - \frac{1}{2\,m}\left[ X''(r) + \frac{2}{r}X'(r)\right] 
	+ V_{\rm eff}(r)\,X(r),
\end{align}
where we defined an effective potential
\begin{align}
 V_{\rm eff}(r) &\equiv \kappa\,(E_0 - m)\,b(r) 
	+ \frac{m}{16}X(r)^2 \nonumber \\
    &= (E_0 - m)\left[\kappa\,b(r) 
	- \frac{1}{8}Y(x)^2\right].
\end{align}
Now considering that $b(r)<0$,  $b'(r)>0$ and $X'(r)<0$ over the whole range of $r$, as shown by the numerical calculations~\cite{ESVW}, the effective potential $V_{\rm eff}$ satisfies 
\begin{equation} 
 3\,(E_0 - m) \lesssim V_{\rm eff}(r)
	\leq 0 
\end{equation}
The quantity $3\,(E_0-m) \gtrsim -0.006\,m$ at $\Delta\lesssim.05$.  This shows that the depth of the effective potential is of $\mathcal{O}(10^{-3}\,m)$ or smaller, implying that produced relativistic axions, which have energy $E\simeq 3 \,m$, can well be regarded as free. The situation would be different in a discussion of dense axion star states, where the binding energy is much more significant; we will return to this topic in a future work.

\end{document}